**Title**

Leveraging Pre-trained Neural Network Models for the Classification of Tumor Cells Analyzed by Label-free Phase Holotomographic Microscopy

**Authors**

Leonor V. C. Losa[#,1], Temple A. Douglas[#,1], Lia Santos[1], Raquel Monteiro[2,3], Isabel Calejo[2,3], Raphaël F. Canadas[2,3], Jana B. Nieder[1,*]

[1] INL – International Iberian Nanotechnology Laboratory, Nieder Group on Ultrafast Bio- and Nanophotonics, Av. Mestre Jose Veiga s/n, 4719-330 Braga, Portugal

[2] Department of Biomedicine, Faculty of Medicine, University of Porto, 4200-450 Porto, Portugal

[3] RISE-Health, Faculty of Medicine, University of Porto, 4200-450 Porto, Portugal

[#] these authors contributed equally to the work
[*] corresponding author: jana.nieder@inl.int

**Abstract**

Can a single label-free image reveal whether cancer cells were exposed to chemotherapy? We present an innovative methodology on the label-free and high-resolution imaging properties of phase holotomographic microscopy coupled with neural network models for the classification of cancer cells. Using 3D phase holotomographic microscopy, we imaged live A549 lung cancer cells with and without paclitaxel, converted stacks to 2D maximum-intensity projections, and evaluated pre-trained convolutional networks (VGG16, ResNet18, DenseNet121, and EfficientNet-B0) for binary classification of treatment status. EfficientNet-B0 achieved 96.9% accuracy on unsegmented images. Refractive index analysis revealed bimodal distribution in treated cells, reflecting heterogeneous biophysical responses to paclitaxel exposure and supporting the network's ability to detect subtle, label-free indicators of drug action. As further proof-of-concept, the same pipeline separated holotomographic images of label-free, high versus low-graded urothelial cancer cells with high accuracy (90.6%). These findings highlight the potential of integrating label-free holotomographic imaging with deep learning techniques for rapid and efficient classification of tumor cells, paving the way for advancements in treatment optimization and personalized diagnostic strategies.

**Introduction**

At present, cancer is among the leading causes of death worldwide, accounting for approximately 9.74 million deaths, and its incidence is expected to increase to an estimated 16.9 million deaths by 2045 [1]. Among the various types of cancer, lung cancer is forecasted to be the leading cause of cancer-related deaths in the world. By 2045, 4.25 million new cases of lung cancer and 3.24 million deaths are expected annually [2]. In this context, understanding the biophysical properties of tumors at a deeper level is essential, particularly in identifying factors that can improve diagnostics and therapeutic outcomes or in optimizing more efficient protocols for diagnostic and treatment optimization [3][4][5].

A major challenge in treating this type of cancer is its heterogeneity in both its treatment response and clinical characteristics, even among tumors of the same pathological type [6][7][8]. This variability makes it challenging to predict an individual's treatment response accurately.

To better address the complexity introduced by tumor heterogeneity, and improve prediction accuracy despite this variability, computational approaches, particularly Machine Learning (ML) have been increasingly applied to cancer research. These algorithms can analyze large datasets of medical images and extract patterns indicative of cellular response to an anticancer treatment, for example [9][10]. Convolutional neural networks (CNNs), including transfer learning with pre-trained models have shown strong results in predicting therapy response [11][12]. For instance, [13] used VGG16 to predict chemotherapy response in non-small cell lung cancer computed tomography (CT) scans with an accuracy of 88.3% with an AUC (area under the receiver operating curve) of 0.982, representing a high model fit to the data.

Even though these approaches have been proven to be successful, these types of images do not capture cellular-level responses to chemotherapy, a step that has been predicted to be an important future methodology for cancer treatment optimization [14]. Thus, an increased interest in live cell imaging techniques has emerged.

Since living cells are naturally transparent, it is challenging to image them with sufficient contrast to observe and analyze their structures. This limitation has been compensated through the use of fluorescent dyes, which enhance visibility but introduce effects such as interference with biological processes, phototoxicity, and photobleaching [15][16].

As an alternative, phase holotomography (PHT) leverages the unique refractive index (RI) properties of cellular components as an intrinsic imaging contrast [17]. This technique enables the visualization of live subcellular features, cellular density and morphology without the need for dyes and fluorescent labels [18][19].

The method relies on the use of a partially coherent laser excitation source that is split into a sample and reference beam via a Mach-Zehnder type interferometric configuration, from which quantitative phase information is reconstructed into a three-dimensional refractive index map [20]. PHT employs low-energy light that minimally perturbs the specimen, reducing phototoxic effects, making it ideal for imaging living cells [21]. Additionally, PHT's numerical refocusing capability enables detailed imaging of cells in three-dimensional volumes [22].

Building upon the capability of PHT systems to provide detailed, label-free insights into cellular responses to therapeutic interventions, this study leverages the combined capabilities of phase holotomographic imaging and a ML pre-trained models of convolutional neural networks to address the challenge of objectively assessing between A549 lung cancer cells treated and untreated with paclitaxel (PTX).

This study represents the first known demonstration of PHT coupled with ML for the purpose of differentiating chemotherapy treated cells from healthy cells. The implementation of this methodology could potentially be utilized in the future to evaluate treatment efficacy in live-cell therapy optimization systems.

## Results

### Phase Holotomography Imaging of A549 cells

Label-free PHT revealed distinct morphological differences between untreated (**Fig. 1I**) and paclitaxel-treated (**Fig. 1II**) A549 cells. Untreated cells displayed heterogeneous adherent morphologies, with well-defined nuclei and vesicles. In contrast, paclitaxel-treated cells displayed a variety of morphologies, including many cells that were rounded and detached, exhibiting typical apoptotic-like features such as membrane blebbing. These morphological changes may be a result of paclitaxel's known method of action in inducing cell death, or other known resistance pathways [23][24].

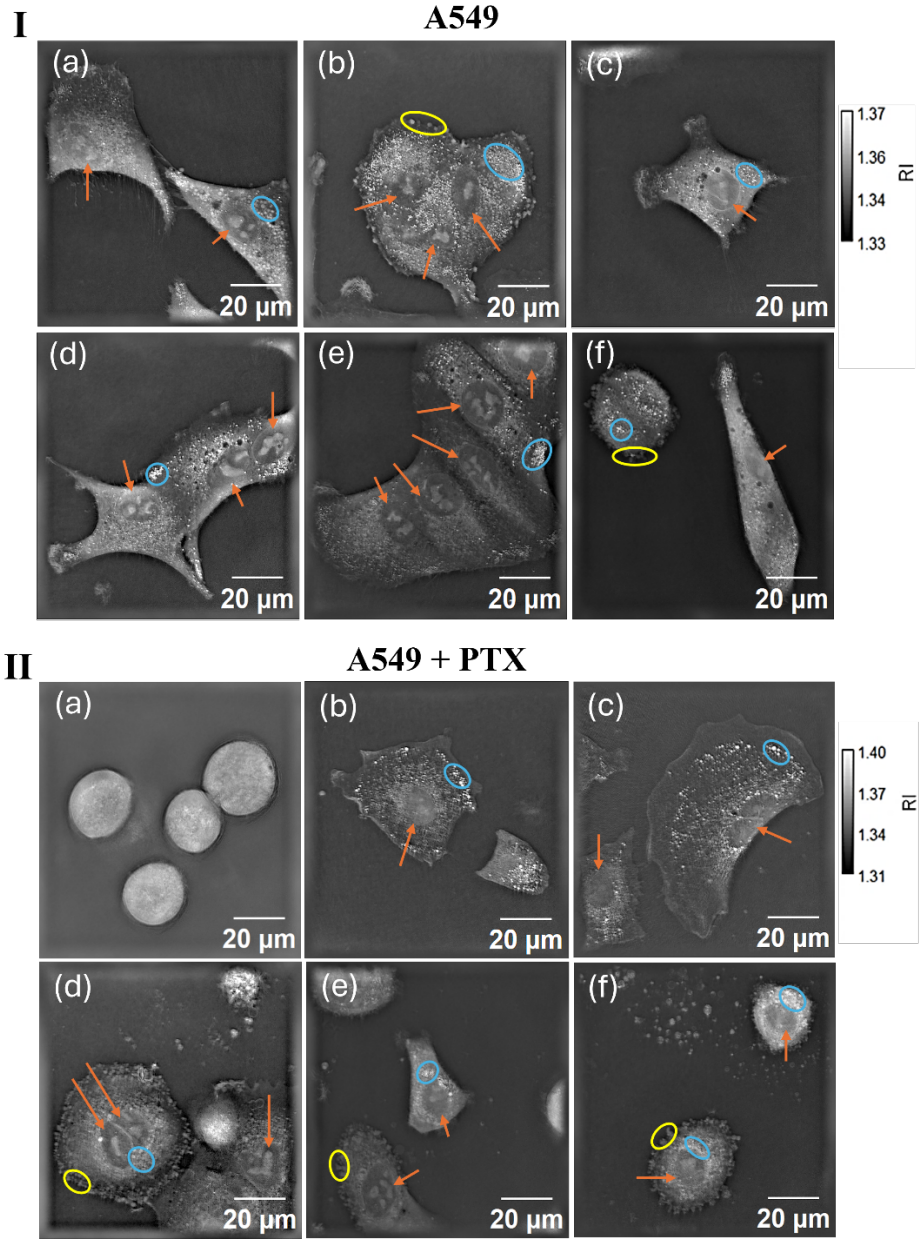

*Fig. 1: Selected phase holotomographic tomography (PHT) images of untreated and treated cancer cells. (I) Untreated A549 cells (a – f), (II) Treated A549 + PTX (a – f). The nucleus is highlighted by orange arrows. Blebs are highlighted by yellow, and vesicles by blue ellipses, respectively.*

Quantitative analysis of the refractive index (RI) distributions calculated for cells segmented using *Cellpose* pipeline to remove background from RI averaging demonstrated that paclitaxel treatment caused a measurable increase in cellular density [25][26]. The mean RI increased from 1.3438 in untreated A549 cells to 1.3539 in treated cells. Each data point in the RI distribution corresponds to the average RI of a single segmented cell, not individual pixels, ensuring that variations reflect cell-level biophysical differences. As shown in **Fig. 2**, the RI distribution also transitioned from unimodal (in untreated A549 cells) to bimodal (in paclitaxel-treated cells). This shift likely arises from the use of the 24h IC50 concentration of paclitaxel, which is cytotoxic, generating a mixed population of viable and apoptotic cells at 24h. The lower RI mode likely corresponds to cells that remain viable and morphologically similar to untreated cells, while the higher RI mode possibly reflects cells undergoing apoptosis or other cell death pathways.

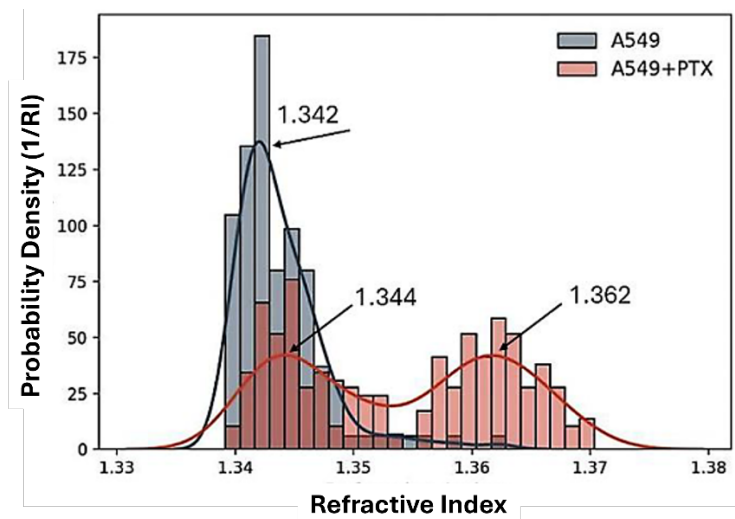

*Fig. 2: Refractive index distribution for A549 cells (gray) and A549 + PTX cells (red), using only segmented cells to avoid averaging with background.*

**Classification of Unsegmented A549 Images**

We trained four pre-trained CNNs (VGG16, ResNet18, DenseNet121, EfficientNet-B0) on maximum-intensity projections of unsegmented PHT images, comprising 359 untreated A549 images and 282 PTX-treated A549 images [27] [28] [29] [30]. Model evaluation was performed using three independent 5-fold cross-validation runs to ensure statistical robustness and mitigate data-split bias [31].

Across all architectures, EfficientNet-B0 consistently outperformed the other models, achieving 96.9 ± 0.6% accuracy, 96.9 ± 0.6% F1-score, and Matthew's correlation coefficient, MCC = 0.94 ± 0.01 (**Table 1**). In comparison, VGG16 and ResNet18 achieved slightly lower MCC values of 0.89 and 0.91, respectively, whereas DenseNet121 approached EfficientNet-B0 with an MCC of 0.93.

*Table 1. Test-set results for selected metrics across 3 runs for different pre-trained models applied to the PTX-treated and untreated A549 data sets.*

| Model | Accuracy ± Std (%) | F1-score ± Std (%) | MCC ± Std |
|---|---|---|---|
| **VGG16** | 94.57 ± 2.19 | 94.59 ± 2.18 | 0.893 ± 0.041 |
| **ResNet18** | 95.35 ± 0.64 | 95.35 ± 0.65 | 0.906 ± 0.014 |
| **DenseNet121** | 96.38 ± 0.37 | 96.39 ± 0.37 | 0.927 ± 0.08 |
| **EfficientNet-B0** | 96.90 ± 0.63 | 96.89 ± 0.64 | 0.938 ± 0.013 |

Because EfficientNet-B0 demonstrated superior performance and stability, we selected it for a more detailed analysis. **Fig. 3** summarizes the validation-stage results obtained during cross-validation for this model only. Panels (A-C) display confusion matrices for the three independent runs. Each matrix shows that the model correctly identified nearly all untreated A549 cells as "A549" (true-negative rate > 97%) and most paclitaxel-treated cells as "A549+PTX" (true-positive rate > 93%). Misclassifications were infrequent (< 4%) and occurred in both classes without a consistent pattern. Panel (D) presents the Receiver Operating Characteristic (ROC) curves for all cross-validation folds of EfficientNet-B0 (across the 3 runs). Each curve lies close to the top-left corner, indicating excellent sensitivity and specificity. The mean Area Under Curve (AUC) was 0.98 ± 0.01, confirming the model's strong discriminative power in distinguishing treated from untreated cells.

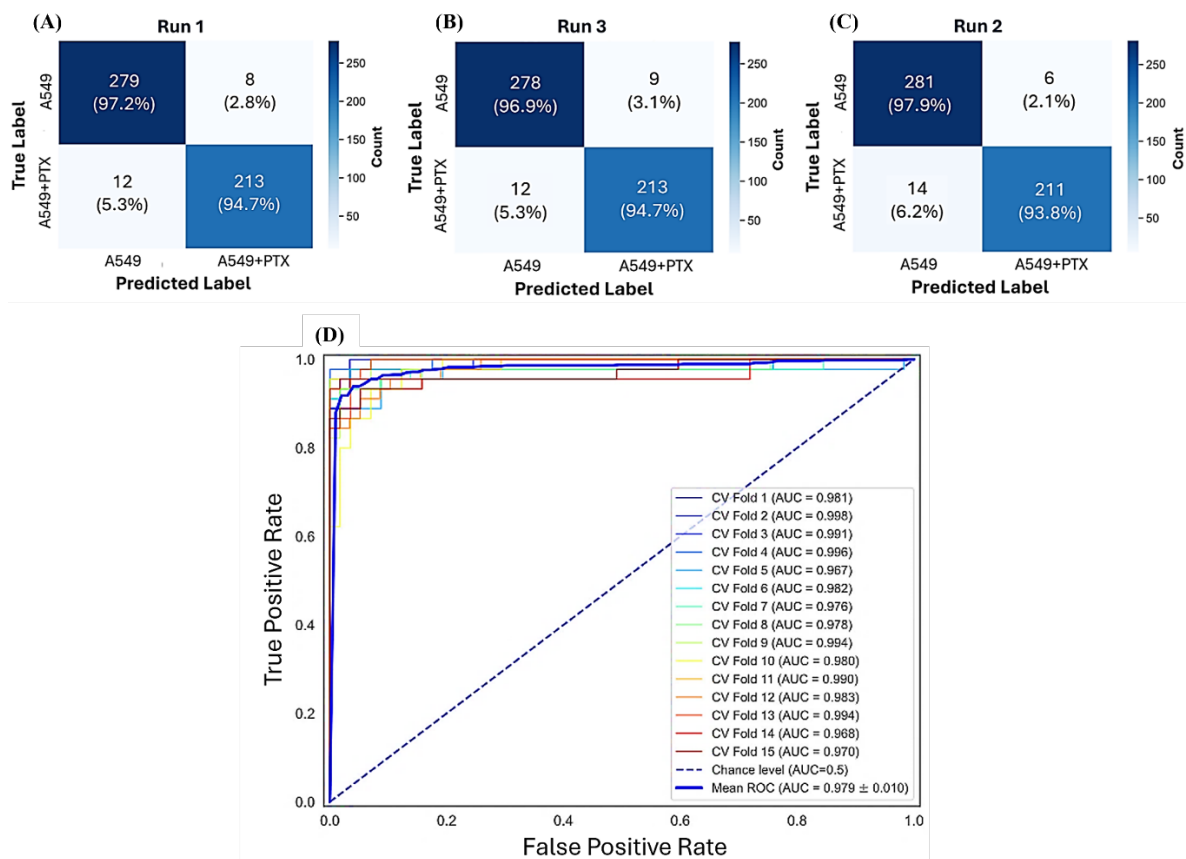

*Fig. 3: Cross-validation confusion matrix for the EfficientNet-B0 model on A549 cells. (A) run 1, (B) run 2, and (C) run 3. (D) Receiver Operating Characteristic (ROC) curve of the EfficientNet-B0 model for the cross-validation (CV) across the 3 runs, for the unsegmented dataset. Folds 1 to 5 are from run 1, folds 6 to 10 from run 2, and folds 11 to 15 from run 3.*

To interpret the mode's internal decision process, we applied Gradient-weighted Class Activation Mapping (Grad-CAM++) [32] to the last convolutional layer of the EfficientNet-B0 model and visualized the resulting attention maps for test-set images (**Fig.4**). In these heatmaps, warmer colors (red/yellow) indicate regions contributing most strongly to the model's prediction, while cooler color (blue/green) represent less relevant areas. For paclitaxel-treated A549 cells (**Fig.4 a-b**), attention was distributed across the cell body, with stronger activations in condensed regions. In contrast, untreated cells (**Fig.4 c-d**) showed attention concentrated along smoother cellular edges and internal areas.

The heatmaps showed that the network's attention was predominantly focused on the cellular regions, particularly within the cell body, rather than on the background. This suggests that the model based decisions on relevant morphological cues instead of artefacts or background noise. Although the highlighted areas did not correspond to specific identifiable subcellular structures, their consistent localization within the cell confirms that EfficientNet-B0 learned to extract meaningful morphological information associated with paclitaxel-induced changes.

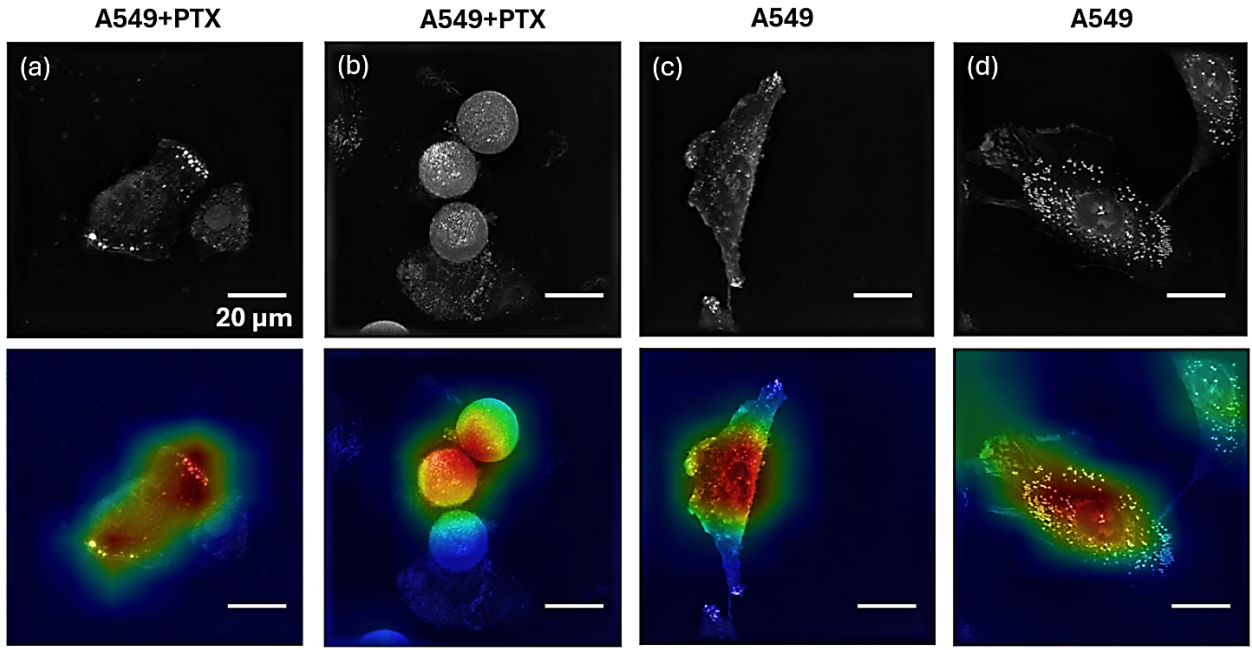

*Fig. 4. Comparison of selected PHT images (top) and heatmaps (bottom) generated with Grad-CAM++ for the EfficientNet-B0 model on A549 and A549 + PTX datasets. These heatmaps were generated for the correctly predicted (a), (b) 'A549+PTX', and (c), (d) 'A549'. All images are from the test-set.*

Together, these findings demonstrate that EfficientNet-B0 reliably distinguishes between untreated and paclitaxel-treated A549 cells from label-free PHT images.

**Validation with Urothelial Cancer Cells**

Tumor grading is a central pathological measure that reflects how closely tumor cells resemble their normal tissue counterparts, providing an indication of malignancy and clinical aggressiveness. Low-grade tumors are composed of cells that are relatively well-differentiated, maintaining organized architecture, uniform nuclei, and slower proliferation rates. In contrast, high-grade tumors contain poorly differentiated cells with marked nuclear atypia, irregular shapes, high mitotic activity, and increased invasiveness [33] [34].

In urothelial carcinoma, grading plays a decisive role in prognosis and treatment strategy. Low-grade tumors are typically non-invasive and managed conservatively, whereas high-grade tumors are more likely to recur or progress and often require aggressive treatment or surveillance. However, grading currently relies on histopathological staining and visual assessment by expert pathologists, a process that is subjective, labor-intensive, and dependent on tissue preparation quality.

The ability to distinguish between tumor grades using label-free, quantitative imaging would offer a valuable alternative for rapid and standardized cancer classification. Because phase holotomographic microscopy measures intrinsic biophysical properties such as cellular refractive index, density, and morphology, it provides contrast directly related to the underlying state of cell differentiation and malignancy. Integrating this information with deep learning models could therefore enable automated, objective grading of tumor cells across cancer types.

To assess the generalizability of our coupled phase holotomographic microscopy and deep learning pipeline, we tested the EfficientNet-B0 pipeline on a different question, to determine if the model could accurately distinguish between low-grade and high-grade nonadherent, paraformaldehyde-fixed, urothelial cancer cells from 10 different sources (8 bladder cancer cell lines divided into 1437 cells from 6 high-grade populations and 473 cells from 2 low-grade populations) (**Fig.5**). These cell lines display broadly similar rounded morphologies typical of non-adherent fixed cells, as they would be found in urine. Because the variations between them are not visually sufficient for reliable discrimination by inspection alone, this highlights the need for quantitative analysis of intrinsic optical properties and ML-based classification.

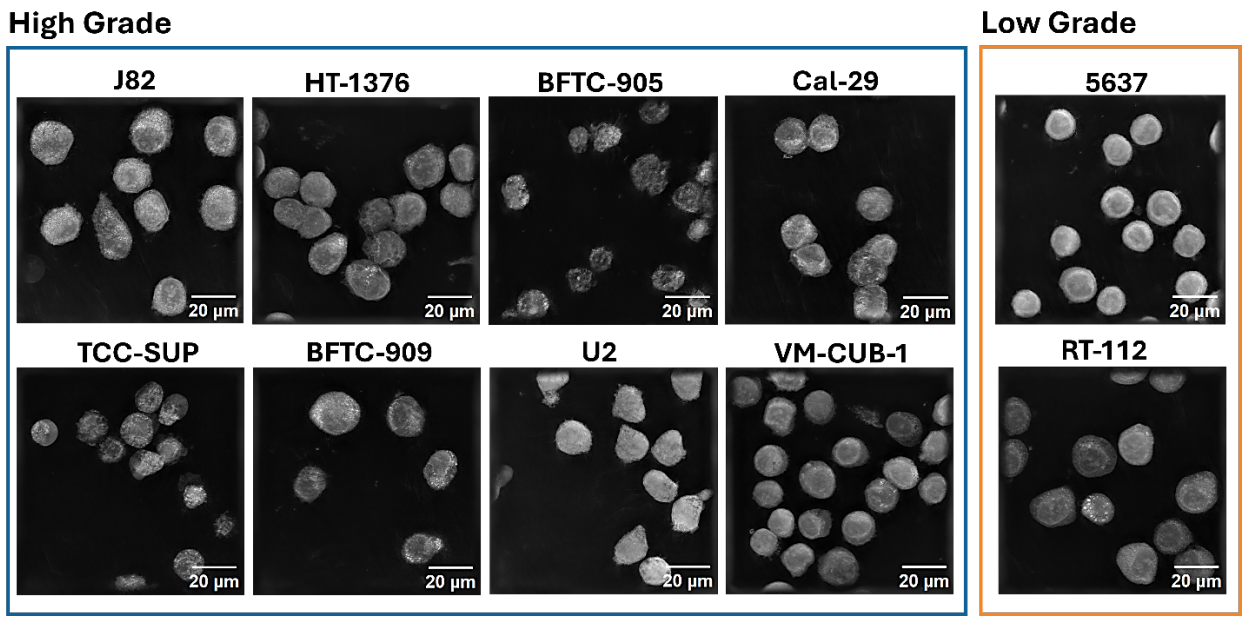

*Fig. 5: Maximum Intensity Projection (MIP) Phase Holotomography Tomography (PHT) images of non-adherent bladder cancer cells. Images on the left represent high grade and, on the right, low grade. The name of the cell line is indicated on top of the image; one image per used cell line is given.*

To investigate whether these subtle differences were associated with measurable biophysical variations, we first analyzed the RI distribution, using only segmented single-cell images. As shown in **Fig.6**, low-grade cells exhibited a narrower RI distribution with a peak at 1.352 and another one at 1.359, while high-grade cells showed broader, also multimodal distributions with additional peaks near 1.341, 1.350, and 1.361. These shifts indicate a possible increased internal heterogeneity and density variations in high-grade cells, consistent with their more disorganized morphology.

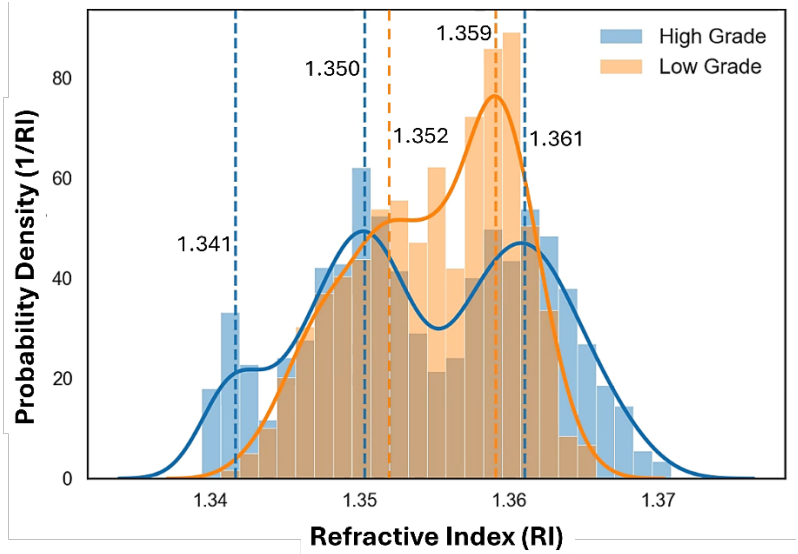

*Fig. 6: Refractive index distribution for bladder cancer cells high-grade (blue) and low-grade (orange), using only segmented cells to avoid averaging with background. The refractive index values for the distributions identified by multi-Gaussian fits are indicated.*

The EfficientNet-B0 model, for this task, was trained in these segmented MIP images using the previously used pipeline and evaluated through three independent runs with different random seeds to confirm result stability. Each run used 5-fold cross-validation, and performance was reported on the independent test set. The validation results across the three runs are summarized in **Fig. 7**, where the confusion matrices (A-C) show consistent classification patterns across repetitions. The model reliably distinguished between high and low-grade cells, with high-grade cells achieving recall values above 94% and low-grade cells achieved recall values above 76% across all runs. The ROC curves for all 15 validation folds (5 folds × 3 runs) are shown in **Fig.7 (D)**, demonstrating strong and stable discriminative performance across all runs, with a mean AUC of 0.94±0.03. The narrow AUC variation confirms the reproducibility of the model pipeline and its robustness.

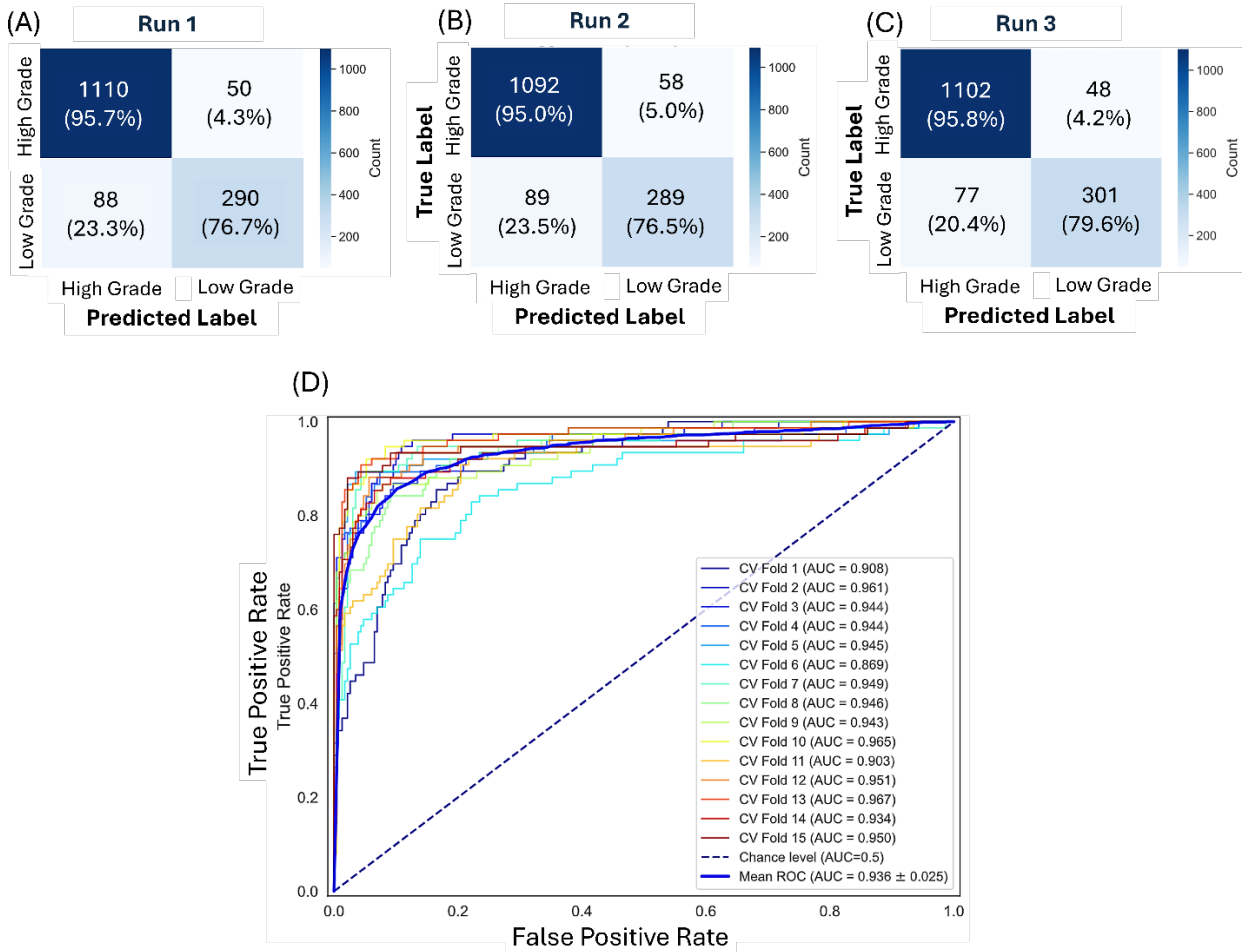

*Fig. 7: Cross-validation confusion matrix for the EfficientNet-B0 model on bladder cancer cells. (A) run 1, (B) run 2, and (C) run 3. (D) Receiver Operating Characteristic (ROC) curve of the EfficientNet-B0 model for the cross-validation (CV) across the 3 runs, for the unsegmented dataset. Folds 1 to 5 are from run 1, folds 6 to 10 from run 2, and folds 11 to 15 from run 3.*

Quantitative metrics for both the single best model and the five-fold ensemble are summarized in **Table 3**. The ensemble achieved 90.58±0.43% accuracy, 90.24±0.47% F1-Score, and a MCC of 0.738±0.013, demonstrating excellent classification performance and high consistency across random initializations.

Together, these results confirm that the proposed label-free deep learning approach can accurately grade urothelial cancer cells based solely on quantitative phase-derived features, without the need for staining or manual extraction. This highlights its potential as an automated, objective, and scalable alternative to conventional tumor grading and possibly broader applications in treatment response assessment.

**Table 3.** *Test-set results for selected metrics across 3 runs for the pre-trained EfficientNet-B0 model applied to high-grade versus low-grade cancer stage urothelial cancer cells (segmented).*

| | Test Set Metrics Urothelial Cancer Cells -Segmented- EfficientNet-B0 | | |
|---|---|---|---|
| | Accuracy (%) | F1-Score (%) | MCC |
| **Single Best Model** | 90.23 ± 0.44 | 90.06 ± 0.46 | 0.733 ± 0.013 |
| **Ensemble** | 90.58 ± 0.43 | 90.24 ± 0.47 | 0.738 ± 0.013 |

## Discussion

Recent advances have been made to push the limits of and resolution of microscopy, using principles of interference and holography to develop high resolution imaging with phase holotomographic microscopy [35]. Recent works have begun posing the question of whether the high-resolution data provided through these novel methods could be coupled with machine learning, with machine learning-influenced grading of pathology samples classified using quantitative phase imaging [36], [37] for better image and morphological reconstruction [38], holotomography for morphological characterization [39], or even for classification of cell types in blood samples or cancers of different organ origin [40].

     Here, we present the first known instance of classification of chemotherapy-treated vs. untreated cells from the same cell population origin using coupled phase holotomographic microscopy with a convolutional neural network-type architecture. In addition, the generalizability of this methodology is exhibited by the application of the model to the grading of a label-free, fixed, nonadherent cell line. We show that pairing label-free live-cell phase holotomography with pre-trained deep learning can classify A549 cells as paclitaxel-treated or untreated with high accuracy. EfficientNet-B0 performed best, and an independent refractive-index shift in treated cells supports the presence of subtle biophysical changes that the model can exploit in characterization. Performance on bladder cancer cells indicates potential portability beyond a single cell type.

     Notably, we can see that the refractive image data tells us that in the treated population of A549 cells and high-grade population of urothelial cancer cells, we see approximately half of the cells centered around a higher refractive index value. In the treated vs. untreated cell case, this could correspond with the fact that at the IC50 value we expect half of the cells to respond to

treatment by undergoing a cell death pathway. High-grade cells exhibit a greater number of mutagenic alterations, leading to a wider spectrum of phenotypic changes compared with low-grade cells. This higher degree of transformation is likely associated with increased heterogeneity in their refractive-index values. However, notably, the machine learning models were able to classify cells at much higher than 50% accuracy (97% for the treated vs. untreated A549 and 91% for the high vs low-grade urothelial cancer cells), indicating that single measurements such as refractive index alone are insufficient for full characterization of these highly heterogeneous cell populations. Similarly, we can expect with these results that these pipelines would also be useful (if provided with a training set for these features) for the classification of different subtypes within these populations.

Phase holotomography alone has been shown to be a valuable tool in the imaging and characterization of sub-cellular features without the use of labels, allowing high-resolution organelle level visualization of live cells [41]. However, there are drawbacks to its implementation in quantitative analysis without the complementary use of machine learning. For instance, holographic reconstructions through thick material regularly overestimate the refractive index of all features inside the cell, and there are known difficulties in deconvolving sample thickness from refractive index, making quantitative RI labeling for the purpose of organelle classification very difficult [42], and features far away from the central plane of the reconstruction are often of lower image quality. In highly rounded or non-adherent cells, features such as organelles, the nucleus, and vesicles are obscured both by not being fixed on a substrate and by the amount of matter that the sample beam must pass through. However, even if these drawbacks impede visual analysis, signatures correlating refractive index with cell depth and clarity of organelles, computationally excluding human bias on the cell features most important for this classification, could surpass and, in fact, utilize the complexity of this information to determine various classifications of stage, treatment resistance, etc.

As this work represents a novel application of the combination of phase holotomography with CNN-type machine learning, we acknowledge important constraints: in-vitro data, a limited number of cell lines and one drug, class imbalance, and segmentation-related variability. Next steps will include expanding datasets across drugs, doses, and cancer types, as well as incorporating 3D models and artificial RI staining to increase data dimensionality. This pipeline should also be tested on greater numbers of clinically relevant samples to assess robustness and real-world utility.

This work demonstrates the implementation of coupled machine learning with phase holotomographic microscopy for the label-free characterization of live adherent lung adenocarcinoma cells treated with chemotherapy as a model for treatment optimization, as well as fixed nonadherent urothelial cancer cells as a model for automatic tumor staging. The selection of these disparate cases and application of the same machine learning coupled to phase holotomography imaging pipeline speaks to the generalizability of this approach. Rapid advances both in the area of machine learning and phase holotomographic microscopy can be expected to improve and advance this technique further. The implementation of this approach could enable rapid, non-destructive readouts for drug testing, treatment optimization, and method development, reducing reliance on stains and complex sample preparation. With appropriate validation, it may contribute to workflows for phenotypic screening and, in the longer term, decision support in personalized therapy research.

## Materials and Methods

### Experimental Design
The objective of this study was to determine whether label-free phase holotomography (PHT) images of lung cancer cells could be used in combination with deep learning to classify paclitaxel-treated versus untreated cells. We designed experiments in three stages: (i) culture and drug treatment of A549 cells, (ii) image acquisition with the PHM system, and (iii) data preprocessing and classification using a pre-trained CNN model. To assess pipeline generalizability, we also tested the optimized pipeline on urothelial cancer cells.

### Cell Line and Culture
Human lung adenocarcinoma A549 cells (ATCC, Virginia, USA) were cultured in Dulbecco's Modified Eagle Medium (DMEM) supplemented with 10% fetal bovine serum (FBS) and 1% penicillin–streptomycin (PS). Cells were maintained at 37ºC in a humidified incubator with 5% $CO_2$ and passaged at ~70% confluence.

Furthermore, we used commercially available urothelial cancer cell lines from high-grade and low-grade patient diagnostics. Cells were cultured in Roswell Park Memorial Institute (RPMI; PAN-Biotech, Germany) 1640 culture medium supplemented with 10% fetal bovine serum (FBS) and 1% penicillin–streptomycin (PS).

Tumor-derived cells were obtained from an ongoing clinical study conducted in collaboration with São João Hospital. All procedures were performed in accordance with the ethical standards of the institutional research committee and the Declaration of Helsinki; ethical approval was granted by the appropriate review board.

Following surgical excision, tumor tissue samples were processed under sterile conditions to remove residual vascular material and obtain clean epithelial tumor fragments. The tissue was mechanically dissociated into small explants and placed in standard cell culture plates. Growth medium was added, and explants were incubated at 37 °C in a humidified 5% $CO_2$ atmosphere. Cell outgrowth and migration from the explants were monitored by phase-contrast microscopy.

Culture supernatant was periodically collected and used for subsequent cell culture steps. Cells were expanded through serial subculturing until a stable adherent cell population displaying consistent proliferation and morphology was established.

Urothelial cancer cells were fixated in suspension using PFA 4% for 30 minutes as a pre-processing step for Phase holotomographic image acquisition.

**Calculation of Paclitaxel IC50**

To determine the half-maximal inhibitory concentration (IC50) of paclitaxel (PTX), $10^4$ A549 cells/well were seeded in 96-well plates and treated with PTX concentrations ranging from 0.0005 to 5 µM. After 24 hours of exposure, cells were washed with PBS and incubated with resazurin (10 µg/mL, Sigma Aldrich) for 4 hours at 37 °C. Fluorescence intensity was measured by exciting at 560 nm and detecting at 590 nm, using a microtiter plate reader (Biotek Synergy H1, Agilent, USA).

**Chemotherapy Drug Treatment**

For imaging, A549 cells were seeded in culture dishes (81218-200, Ibidi, Germany) chambers and allowed to adhere for 24 hours at 37 °C with 5% $CO_2$. After this period, the culture medium was replaced with fresh medium containing 5.698 µM PTX, corresponding to the $IC_{50}$ for this cell line. After 24h in the medium with PTX, we took the cells for imaging.

**Phase Holotomographic Image Acquisition**

Three-dimensional phase holotomographic images were acquired using a phase holotomographic microscope (3D Cell Explorer, Nanolive S.A., Switzerland) that splits a semi-coherent laser (nm) into a Mach-Zehnder configuration where one arm passes through the sample and the other Is used as a reference for reconstructing a digital hologram.

The microscope utilized a dry 60×/N.A. 0.8 objective and a CMOS camera (IMX174 CMOS, Sony, Korea) camera. The system operates with a class 1 laser (λ=520 nm, 0.2 mW/ $mm^2$).

Interference between the reference and sample beams generated an interference pattern, which was then transformed by the STEVE software (Nanolive S.A., Switzerland) into 96 2D holograms, subsequently reconstructed into 3D tomograms.

Environmental controls during imaging were rigorously maintained. 2 hours prior to each experiment the gas mixer was set to 0.5 L/min for air and 0.02 L/min for CO2, and the temperature controller was set to 37 °C. An air pump was activated, and the two chambers within the incubator were filled with distilled water to ensure high humidity and prevent sample evaporation. The maximum field of view for image acquisition was $90 \times 90 \times 30$ $\mu m^3$.

**Data Preprocessing**

To prepare volumetric data for analysis, we first trimmed non-informative slices from the top and bottom of each 3D image stack, removing sections that contained only background signal or reconstruction noise and no visible cellular content. This ensured that the subsequent projections represented only the biologically relevant regions of the sample. After trimming, maximum intensity projection (MIP) images were generated to obtain two-dimensional representations of the 3D datasets. Each original stack consisted of 96 sliced, and each projected image had a size of 512 ×512 pixels, corresponding to an imaging area of approximately 90×90 µm.

For the ML pipeline, all images were resized to 224 ×224 pixels to match the input size required by the pre-trained convolutional neural network architectures used in this study.

**Image Segmentation**

For CNN classification of the urothelial cancer cell lines and RI calculation of all cell lines, we used *Cellpose* to segment multi-cell images into individual cells. The segmentation process produced binary masks, where each detected cell was assigned a unique label corresponding to its outline. These masks were used to extract individual cells from the MIP images, generating cropped single-cell images. Because the segmented cells varied in size and shape, each cropped image was centered on a black background and padded with black pixels to create uniform 224×224 pixel frames suitable for input into the ML models.

Segmentation performance varied according to cell type. For A549 cells, which were imaged live, segmentation was limited by frequent cell overlapping and boundary ambiguity, leading to the exclusion of some images from the segmented dataset. For this reason, segmented images were only used for RI analysis and not for CNN classification. In contrast, urothelial cancer cells were chemically fixed before imaging, resulting in clearer boundaries and more stable morphology, which enabled successful segmentation of all imaged cells.

**Deep Learning Pipeline**

We adapted four ImageNet pre-trained convolutional neural networks (EfficientNet-B0, DenseNet121, ResNet18, and VGG16) for grayscale PHT images. Since our images are single-channel, the original 3-channel input convolutions were replaced by single-channel versions initialized from a normal distribution ($\mu=0$, $\sigma=0.001$). All other layers retained pre-trained weights. All models were originally trained on ImageNet (1000 classes, RGB input), so their final fully connected classifier layers were replaced with 2-unit output layers corresponding to the two study classes (untreated and PTX-treated A549 cells, and high-grade vs low grade).

For EfficientNet-B0, the output of the average pooling layer (1280 features) was passed to a fully connected layer of 640 units followed by dropout. A second dropout was added before the final classification layer (2 output units).

For DenseNet121, we inserted a dropout layer between the 1024-dimensional pooled features and the final classifier, also changed to 2 output units.

For VGG16, the two existing dropout layers were retained, but their probabilities were optimized instead of fixed at 0.5, and the output layer was replaced by a 2-unit classifier.

For ResNet18, we added a dropout layer between the 512-dimensional pooled features and the output layer (2 units).

All dropout probabilities were tuned using *Optuna's Bayesian* optimization (TPESampler) within the range [0.25-0.48] for the first dropout and [0.18-0.55] for the second. *Optuna* also optimized the learning rate ([0.0002-0.0015]) and weight decay ([0.0005-0.05]), targeting the Matthews Correlation Coefficient (MCC) as the objective metric.

Training used 5-fold stratified cross-validation with early stopping (patience=15) and data augmentation (random flips and rotations) applied only to training subsets. Images were resized to 224x224 pixels and z-score normalized using training statistics per fold to precent data leakage.

To ensure robustness against initialization variance, all experiments were repeated three times with different random seeds (53, 65 and 88), and metrics were averaged across the three runs. Performance was reported for both the single best model (highest validation accuracy) and an ensemble approach. In the ensemble, multiple independently trained models contributed predictions for each test sample, and the averaged *softmax* probabilities were used to determine the final class output. This strategy reduces the impact of variance from random initialization and training splits, improving overall prediction stability.

**Acknowledgments**

This research was supported by the "La Caixa" Foundation (ID 100010434) and FCT, I.P. via the project Diamond4Brain under the agreement LCF/PR/HP20/5230000, by the Health from Portugal project (C630926586-00465198) supported by Component C5 – Capitalisation and Business Innovation, under the Portuguese Resilience and Recovery Plan, through the NextGenerationEU Fund, by the CELLo Project (2022.05237.PTDC) funded by FCT; the CELLECTED Project (Grant no.: CI24-10322) supported by "la Caixa" Foundation via the CaixaImpulse Program. The authors also thank the Urology Service of São João Hospital for their collaboration and the Nanophotonics and Bioimaging Facility (NBI) at the International Iberian Nanotechnology Laboratory (INL) for support, especially Dr. Mariana Carvalho for training and support for the 3D Cell explorer assays.

[34] Y. Yu and M. R. Downes, "Papillary Urothelial Neoplasms: Clinical, Histologic, and Prognostic Features," in *Urologic Cancers*, N. Barber and A. Ali, Eds. Brisbane (AU): Exon Publications, 2022.

[35] V. Chvalova, T. Vomastek, and T. Grousl, "Comparison of holotomographic microscopy and coherence-controlled holographic microscopy," *J. Microsc.*, vol. 294, no. 1, pp. 5–13, 2024, doi: 10.1111/jmi.13260.

[36] T. H. Nguyen *et al.*, "Automatic Gleason grading of prostate cancer using quantitative phase imaging and machine learning," *J. Biomed. Opt.*, vol. 22, no. 3, p. 036015, 2017, doi: 10.1117/1.jbo.22.3.036015.

[37] D. Roitshtain, L. Wolbromsky, E. Bal, H. Greenspan, L. L. Satterwhite, and N. T. Shaked, "Quantitative phase microscopy spatial signatures of cancer cells," *Cytom. Part A*, vol. 91, no. 5, pp. 482–493, 2017, doi: 10.1002/cyto.a.23100.

[38] J. Kim, Y. Kim, H. S. Lee, E. Seo, and S. J. Lee, "Single-shot reconstruction of three-dimensional morphology of biological cells in digital holographic microscopy using a physics-driven neural network," *Nat. Commun.*, vol. 16, 2025, doi: 10.1038/s41467-025-61374-0.

[39] V. K. Lam, T. C. Nguyen, B. M. Chung, G. Nehmetallah, and C. B. Raub, "Quantitative assessment of cancer cell morphology and motility using telecentric digital holographic microscopy and machine learning," *Cytom. Part A*, vol. 93, no. 3, pp. 334–345, 2018, doi: 10.1002/cyto.a.23316.

[40] K. Jaferzadeh, S. Son, A. Rehman, S. Park, and I. Moon, "Automated Stain-Free Holographic Image-Based Phenotypic Classification of Elliptical Cancer Cells," *Adv. Photonics Res.*, vol. 4, no. 1, 2023, doi: 10.1002/adpr.202200043.

[41] K. Bubb *et al.*, "Metabolic rewiring caused by mitochondrial dysfunction promotes mTORC1-dependent skeletal aging," *Sci. Adv.*, vol. 11, no. 16, 2025, doi: 10.1126/sciadv.ads1842.

[42] G. Dardikman and N. T. Shaked, "Review on methods of solving the refractive index–thickness coupling problem in digital holographic microscopy of biological cells," *Opt. Commun.*, vol. 422, no. April, pp. 8–16, 2018, doi: 10.1016/j.optcom.2017.11.084.
Page **20** of 20